\documentclass[manuscript]{aastex}
\usepackage{emulateapj5,epsfig,graphicx,ifthen}
\usepackage{natbib}

\newcommand{\figtype}{EPS}
\def\myputfigure#1#2#3#4#5%
{\vskip#5pt\makebox[0pt]{\hskip#2in
\includegraphics[width=#3\textwidth]{#1}}\vskip#4pt\hfill}

\newlength{\colwidth}

\def\plottwr#1#2{\centering \leavevmode
\epsfysize=.42\textwidth \epsfxsize=.45\textwidth \epsfbox{#1} \medskip
\epsfysize=.33\textwidth \epsfxsize=.45\textwidth \epsfbox{#2}}

\def\plottws#1#2{\centering \leavevmode
\epsfysize=.33\textwidth \epsfxsize=.45\textwidth \epsfbox{#1} \medskip
\epsfysize=.33\textwidth \epsfxsize=.45\textwidth \epsfbox{#2}}

\newenvironment{inlinefigure}{
\def\@captype{figure}
\medskip\noindent\begin{minipage}{0.999\linewidth}\begin{center}}
{\end{center}\end{minipage}\medskip}

\shortauthors{KEMPNER, SARAZIN \& RICKER}
\shorttitle{{\it CHANDRA} OBSERVATION OF ABELL 85}

\submitted{Accepted for publication in the Astrophysical Journal.}

\begin{document}

\title{Chandra Observations of Abell 85: Merger of the South Subcluster}

\author{Joshua C. Kempner and Craig L. Sarazin}
\affil{Department of Astronomy, University of Virginia,
P. O. Box 3818, Charlottesville, VA 22903-0818}
\email{jck7k@virginia.edu,
cls7i@virginia.edu}

\and\author{Paul M. Ricker}
\affil{Department of Astronomy \& Astrophysics, University of Chicago,
5640 S.\ Ellis Ave., Chicago, IL 60637}
\email{ricker@flash.uchicago.edu}

\begin{abstract}
We present an analysis of a highly asymmetric cluster merger from a {\it
Chandra} observation of Abell~85.
The merger shows significant disruption of the less massive subcluster
from ram pressure effects.
Nevertheless, a cold core, coincident with the cD galaxy, is observed to
persist in the subcluster.
We derive dynamical information from the motion of the cold core through
the main cluster's ICM.
Multiple derivations of the velocity of the core suggest a Mach number
of $\mathcal{M} \approx 1.4$, or $v \sim 2150$~km~s$^{-1}$, though with
substantial uncertainty.
We construct a consistent kinematic model for the merger based on this
dynamical analysis.
As has been found for other such ``cold fronts,'' conduction appears to
be suppressed across the front.
Thermal conduction may be suppressed by a magnetic field with a
significant component perpendicular to the subcluster's direction of
motion.
The effect of the merger interaction in creating and shaping the
observed radio sources is also discussed.
It appears most likely that the radio source is due to distorted
and detached lobes from the subcluster cD galaxy, rather than being
a radio halo.
\end{abstract}


\section{Introduction}
\label{sec:south_intro}

Mergers of clusters of galaxies are highly energetic events, releasing
a total kinetic energy of $\sim 10^{63}$~ergs into the intracluster
medium (ICM).
When clusters merge, shocks are driven into the ICM, dissipating the
kinetic energy of the merger and heating the gas.
These shocks also have nonthermal effects, including the generation of
turbulence in the ICM and acceleration of charged particles to
relativistic, or cosmic ray, energies.
Observations with {\it Chandra} of merging clusters have provided new
insights into the cluster merger process, including the unpredicted
discovery of the persistence of cold cores from pre-merger cooling flows
well into the lifetime of a merger
\citep[``cold fronts:''][]{mpn+00,vmm01b}.

Abell~85 is in the early stages of merging with two subclusters, each
much less massive than the main cluster.
One subcluster is merging from the southwest while the other subcluster
is merging from the south.
The south subcluster will be the focus of our discussion here, while the
other subcluster and its associated radio relic will be discussed in a
later paper.
Abell~85 is unusual in being one of the few clusters known to be in the
process of a merger while maintaining a moderate
\citetext{$107~M_\odot$~yr$^{-1}$; \citealp{pfe+98}}
cooling flow.
Presumably, this implies that the merging subclusters have not yet penetrated
the inner few hundred kiloparsecs of the cluster and have therefore not
yet been able to disrupt the cooling flow.

The south subcluster is more massive than
the southwest subcluster.
There has been some uncertainty in the past as to whether or not this
subcluster is in fact merging with the main cluster or is merely seen
against the main cluster in projection.
Using data from {\it ASCA},
\citet{mfs+98} determined that the temperature in the region of the
subcluster is the same as or slightly greater than that of the rest of
the main cluster at the same radius.
If the subcluster were not merging and were only seen it projection,
its smaller mass would give it a lower temperature than that of the main
cluster.
Thus, the higher temperature
indicates that the subcluster is almost certainly interacting.

The redshifts of the galaxies in the southern subcluster are slightly
larger that those of the main cluster
\citep{bfh+91,dfl+98}.
This suggests that the southern subcluster is either a background cluster
or that it is slightly in front of the main cluster and its excess redshift
comes from its peculiar motion as it falls into the main cluster.
Based on the analysis of \citet{mfs+98} and the observations presented in
the present paper, we believe that it is merging with the main cluster.
Thus, we will assume that the southern subcluster is at essentially
the same distance as the main cluster, and that any difference in their
observed redshifts is caused by their relative motion along the line of
sight as they merge.

The {\it Chandra} observation and basic data reduction are discussed in
\S~\ref{sec:south_data}.
The X-ray image is presented in \S~\ref{sec:south_xray_image}.
In \S~\ref{sec:south_xray_spectra}, we analyze the spectra of interesting
regions associated with the southern subcluster.
The profiles of the X-ray surface brightness and temperature within the
subcluster and in the region ahead of the subcluster are extracted in
\S~\ref{sec:south_xray_profiles}.
We discuss the evidence for a merger and X-ray determinations of the
merger Mach number in \S~\ref{sec:south_hydro}.
The pressure increase at the cold front and properties of the bow shock
are used to derive the merger velocity in \S~\ref{sec:south_hydro_stag}
and \ref{sec:south_hydro_bow}.
We construct a consistent kinematic model for the merger in
\S~\ref{sec:south_kine}.
The suppression of conduction across the cold front is discussed
briefly in \S~\ref{sec:south_trans_conduction}.
There have been claims of a possible radio relic in this cluster as well
\citep{bpl98}, which we discuss in \S~\ref{sec:south_radio}.
Our results are summarized in \S~\ref{sec:south_summary}.
We assume $H_0 = 50$~km~s$^{-1}$~Mpc$^{-1}$ and $q_0 = 0.5$ throughout
this paper.
At the cluster redshift of $z = 0.0538$, 1\arcsec\ corresponds to 1.43~kpc.
All of the errors quoted are at the 90\% confidence level.

\section{Observation and Data Reduction}
\label{sec:south_data}

Abell 85 was observed with ACIS-I detector on {\it Chandra} in a single
39~ksec observation.
Using the count rate in the S3 chip,
we excluded data during two small background flares using the
{\it lc\_clean}\footnote{see
\url{http://hea-www.harvard.edu/$\sim$maxim/axaf/acisbg/}} routine
written by
Maxim Markevitch.
This left 36,587~s of useful exposure time.
Although the observation included the four ACIS-I chips and the
S3 and S4 chips, the analysis presented here will be based on
the ACIS-I data only.
The focal plane temperature during the observation was $-$120~C.
A raw image of the entire ACIS-I detector in the spectral band 0.3--10
keV is presented in Figure~\ref{fig:south_image}.

\begin{inlinefigure}
   \ifthenelse{\equal{\figtype}{EPS}}{
     \epsfxsize=.45\textwidth
     \epsfysize=.45\textwidth
     \centerline{\epsfysize=0.83\colwidth\epsfxsize=\colwidth\epsffile{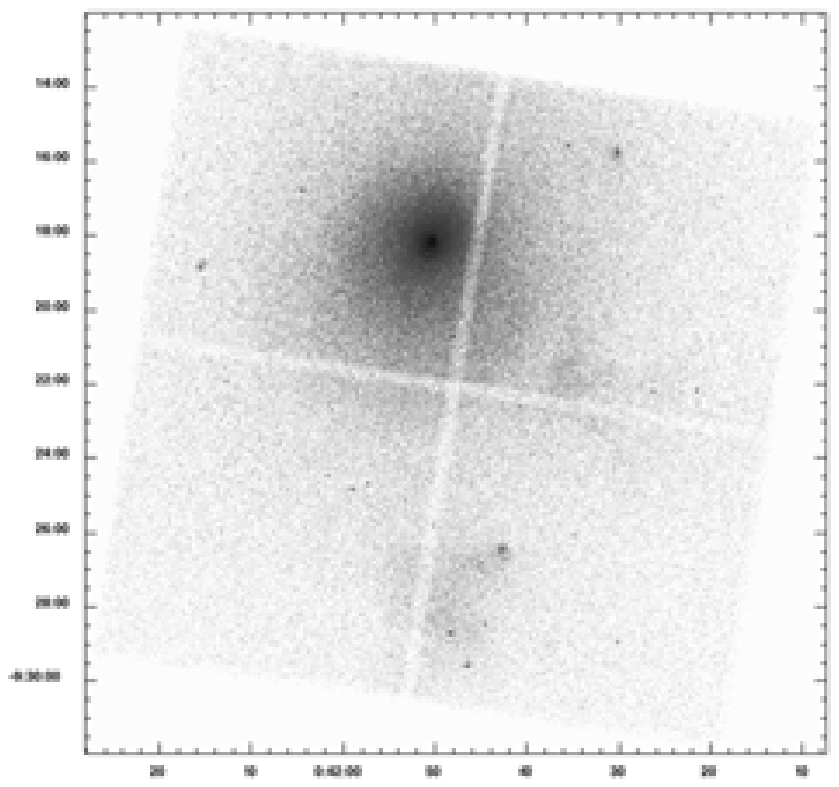}}
  }
  {
  }
\figcaption[f1.eps]{Raw X-ray image of Abell~85 in the 0.3--10 keV band,
uncorrected for background or exposure.
All four ACIS-I chips are shown; the regions of reduced exposure are the
interchip gaps.
The center of the main cluster and cooling flow are located on the upper
left chip,
the center of the southwest subcluster is located just above the lower edge of
the upper right chip,
and the south subcluster is near the bottom and overlaps the two lower chips.
\label{fig:south_image}}
\end{inlinefigure}

The ACIS-I suffers from enhanced charge transfer inefficiency (CTI)
caused by radiation damage early in the mission.
We corrected for the quantum efficiency non-uniformity and gain variations
caused by this damage,
but the degradation in spectral response was not corrected for due to the
lack of availability of appropriate response matrices.
We used version~1 of the January 29, 2001 gain file and response files.
Because of uncertainties in the spectral response at low energy,
we limit our spectral analyses to the range 0.7--10.0~keV, excluding the
1.8--2.2~keV band around the mirror iridium edge.
We constructed blank sky backgrounds from the March 23, 2001 versions of
Maxim Markevitch's ACIS-I background photon lists using the routine
{\it make\_acisbg}\footnotemark[1].

\section{X-ray Properties of the Southern Subcluster}
\label{sec:south_xray}

\subsection{X-ray Image}
\label{sec:south_xray_image}

Figure~\ref{fig:south_smooth_image} shows an adaptively smoothed image
of an approximately 6\farcm4$\times$6\farcm4 region around the southern
subcluster.
The image was smoothed to a signal-to-noise ratio of three for each
smoothing beam.
The same set of smoothing kernels were used to smooth the blank-sky
background image and the exposure map.
The smoothed background image was subtracted from the smoothed
subcluster image,
and the result was divided by the smoothed exposure map.

\begin{inlinefigure}
   \ifthenelse{\equal{\figtype}{EPS}}{
     \epsfxsize=.45\textwidth
     \epsfysize=.45\textwidth
     \centerline{\epsfysize=0.89\colwidth\epsfxsize=\colwidth\epsffile{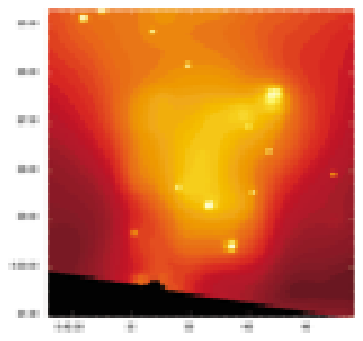}}
  }
  {
  }
\figcaption[f2.eps]{Adaptively smoothed image of the south subcluster,
corrected for background and exposure.
The cold core is at the NW corner of the subcluster.
\label{fig:south_smooth_image}}
\end{inlinefigure}

The subcluster is roughly cone-shaped (Figure~\ref{fig:south_smooth_image}),
with a high surface brightness knot at the northwest corner.
This bright region is spatially extended, with a diameter of about
26\arcsec,
and is centered on the cD galaxy which is the brightest galaxy in
the southern subcluster
(see Figure~\ref{fig:south_radio} below).
The northern edge of this knot shows an abrupt surface brightness edge.
This suggests that it is either a merger shock or a ``cold front,'' the
leading edge of a cold core.
There is a curved tail of brighter X-ray emission extending to the
southeast of the bright knot.

\begin{table*}[ht]
\begin{center}
\footnotesize
\caption{Spectral fit parameters \label{tab:south_spec_fits}}
\begin{tabular}{lcccccc}
\tableline
\tableline
 & & $k_B T$ & $Z$ & & & \\
Region & model & (keV) & ($Z_\odot$) & $\chi^2$ & $d.o.f.$ & net counts \\
\tableline
cold core & mekal   & $2.3^{+0.6}_{-0.4}$ & $0.48^{+0.97}_{-0.27}$ & \phn16.7 & \phn16 & \phn416 \\
cold core & mekal $+$ main cluster & $2.1^{+0.5}_{-0.4}$ & $0.53^{+0.65}_{-0.34}$ & \phn15.9 & \phn16 & \phn416 \\
subcluster $-$ cold core& mekal   & $6.3^{+0.6}_{-0.5}$ & $0.34^{+0.15}_{-0.15}$ & 254.0 & 196 & 8486 \\
subcluster $-$ cold core& mekal $+$ main cluster & $5.5^{+0.7}_{-0.6}$ & $0.36^{+0.20}_{-0.19}$ & 254.6 & 196 & 8486 \\
main cluster& mekal   & $9.0^{+2.4}_{-1.7}$ & $0.52^{+0.67}_{-0.50}$ & 186.8 & 168 & 4527 \\
\tableline
\end{tabular}
\end{center}
\end{table*}

The overall geometry indicates that the ICM in the subcluster has
been affected by ram pressure from the gas in the main cluster.
The morphology suggests that the south subcluster is in the early
stages of merging with the main cluster, and that it is falling into
the main cluster for the first time from the south.
The sharp edge at the top of the bright knot is symmetrical about a
position angle of $-15^\circ$ ($15^\circ$ west of north).
The bulk of the subcluster forms a conic distribution centered about
a position angle of $123^\circ$ ($58^\circ$ east of south).
This difference between $123^\circ$ and
$165^\circ = ( 180^\circ - 15^\circ )$
may indicate that the cD galaxy is moving relative
to the center of mass of the subcluster, or that the shape has been
affected by the density structure in the subcluster and main cluster gas.
In any case, it appears that the transverse component of the
velocity of the subcluster relative to the main cluster lies at
a position angle between $-60^\circ$ and $-10^\circ$.
The position angle of the center of the main cluster from the subcluster is
about $+13^\circ$.
This implies that this is an offset merger;
the collision is occurring with a nonzero impact parameter and angular
momentum.

Several point X-ray sources are also seen in the region of the subcluster
(Figure~\ref{fig:south_smooth_image}).
Only two of these have a an optical counter part:
the source at
R.A.~=~$00^{\rm h}41^{\rm m}50\fs4$, Dec.~=~$-9\arcdeg25\arcmin48\arcsec$
is coincident with the nucleus of the galaxy PGC~93226, which is
a cluster member and a radio source
(radio source B in Figure~\ref{fig:south_radio} below);
and the source at
R.A.~=~$00^{\rm h}41^{\rm m}59\fs0$, Dec.~=~$-9\arcdeg24\arcmin49\arcsec$
is coincident with a very faint galaxy
of unknown redshift
\citetext{source 1-2117; \citealp{sdg+98}}.

\begin{figure*}[ht]
\plottwr{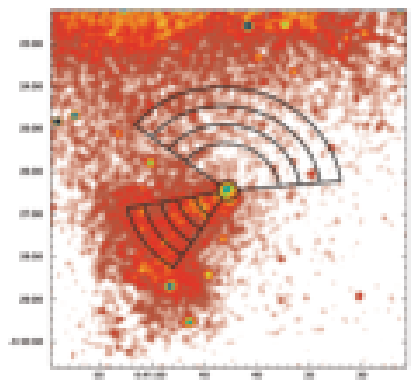}{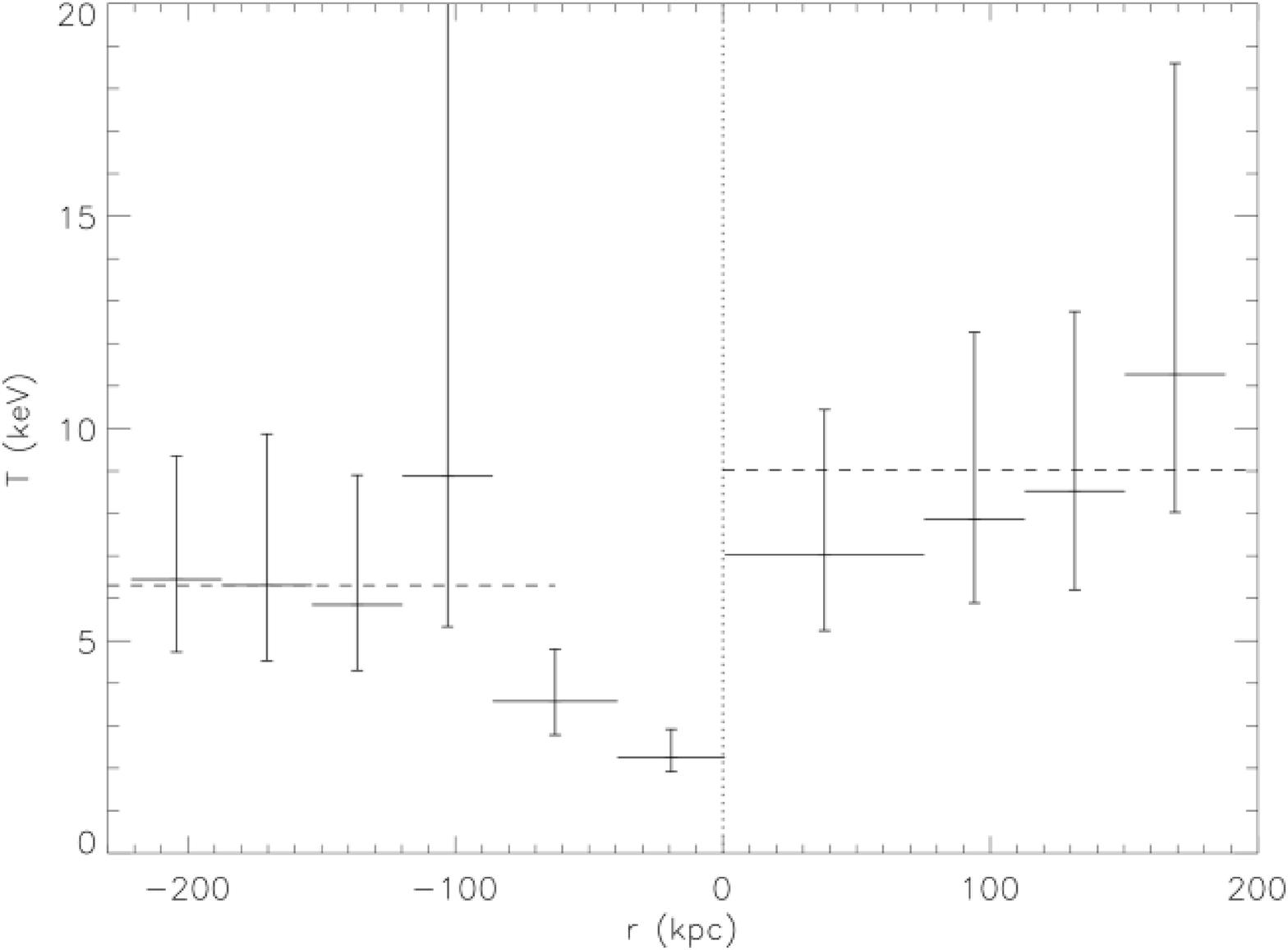}
\plottws{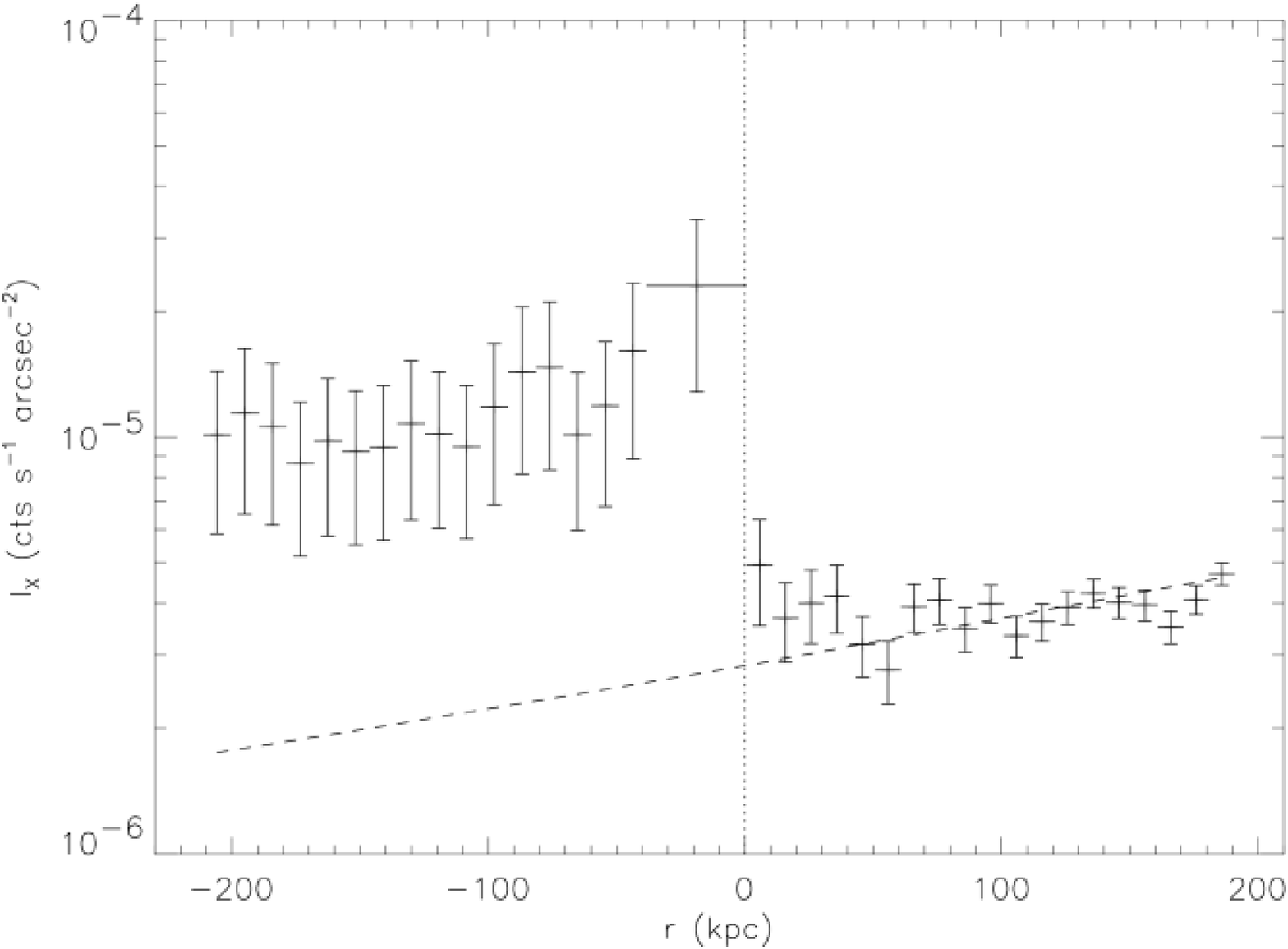}{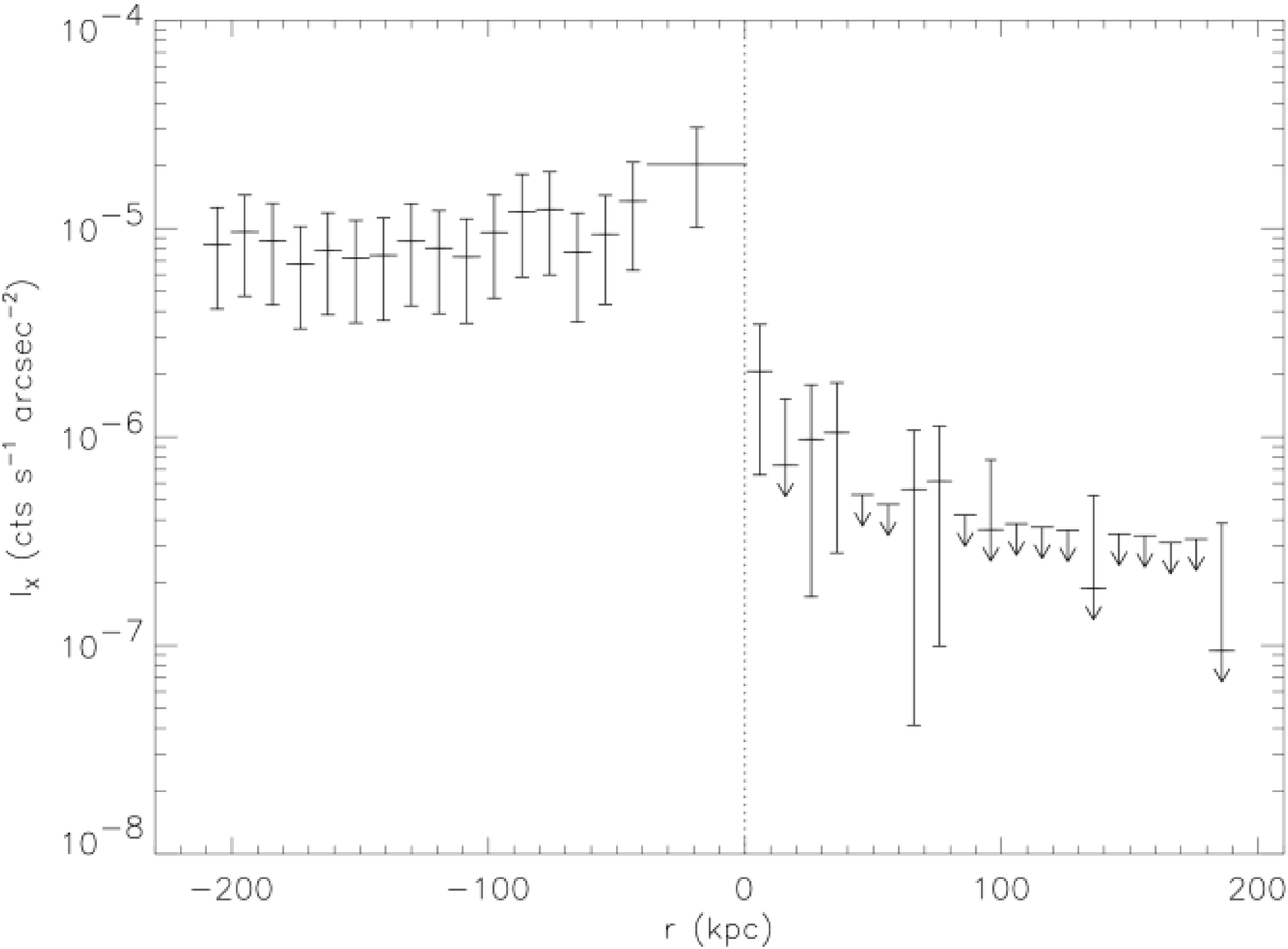}
\figcaption[f3a.eps,f3b.eps,f3c.eps,f3d.eps]{{\it (a)} Elliptical
annular wedge regions used for determining temperatures are shown
superposed on a gaussian smoothed image of the south subcluster.
{\it (b)} Temperature profile in the regions shown in {\it (a)}.
The dotted line indicates the position of the cold front, with positive radii
in the direction of motion of the subcluster (i.e., ahead of the cold
front).
The dashed lines indicate the temperatures from the single-temperature
fits given in Table~\ref{tab:south_spec_fits}.
The fit to the subcluster is shown for the radii over which it was
determined.
{\it (c)} X-ray surface brightness from 0.3-10~keV in a set of elliptical
annular wedges with the same shape and orientation as those in {\it (a)},
but more closely spaced.
{\it (d)} X-ray surface brightness values from {\it (c)}, after correction
for the foreground and background emission from the main cluster.
The dashed curve in {\it (c)} is the projected surface brightness
contribution from the main cluster, which is used to correct the values
in {\it (d)}.
\label{fig:south_prof}}
\end{figure*}

\subsection{X-ray Spectra}
\label{sec:south_xray_spectra}

We extracted the X-ray spectra of the bright knot coincident with the
cD galaxy at the top of the subcluster (``cold core''),
of the remainder of the subcluster,
and of the main cluster gas at a similar projected distance from
the center of the main cluster.
The spectrum of the bright knot was extracted from the elliptical region
at the center in Figure~\ref{fig:south_prof}a.
The spectrum from the remainder of the subcluster was extracted from a
polygonal region encompassing most of the rest of the subcluster, minus
the point sources within that region.
The main cluster spectrum came from an annular pie wedge to the east of
the subcluster and which encompassed the same radii from the cluster
center as the subcluster.
The spectral fits in these regions are presented in
Table~\ref{tab:south_spec_fits}.
The spectra were grouped to have a minimum of 20 counts per channel.
As described in \S~\ref{sec:south_data}, we restricted our spectral
analysis to the range 0.7--10.0~keV, minus the band from 1.8--2.2~keV.
In the fit to the spectrum of the main cluster, we cut the spectrum off
at 9.0~keV because the spectrum of this diffuse emission is dominated
by background above this energy.
The spectra were fit within {\it xspec} using the {\it mekal} model for
the thermal emission.
The absorption column was fixed at the Galactic value of $2.85 \times
10^{20}$~cm$^{-2}$ \citep{dl90}.
Since the observed emission from the cold core and subcluster presumably
contain emission from the main cluster seen in projection in front of
and behind these regions, we also fit the spectrum of the cold core and
subcluster with two {\it mekal} thermal components, with the
shape of the hotter component fixed at the values found for the main
cluster, and the normalization determined by the relative areas of the
regions.

We find that the temperature of the bright knot at the top of the
subcluster ($2.1^{+0.5}_{-0.4}$~keV) is much lower than the temperature
of the remainder of the subcluster or of the surrounding gas from
the main cluster.
This shows that this X-ray bright and dense region is a cooling core
associated with the central region and central cD galaxy in the
subcluster.
The sharp surface brightness discontinuity at the northern edge of this
knot must be a cold front, rather than a merger shock, since the compressed
gas has a lower temperature and specific entropy than the less dense gas
\citep{mpn+00,vmm01b}.

We also fit the cold core with a cooling flow model, with and without an
additional foreground and background contribution from the main cluster.
In both cases, the gas was allowed to cool to the minimum allowable
temperature, essentially zero.
For the fit without the additional component for the main cluster,
we fixed the maximum temperature and abundance to those from the third
fit in Table~\ref{tab:south_spec_fits}.
We found a cooling rate of $7.3^{+0.7}_{-0.8}~M_\odot$~yr$^{-1}$.
With an additional mekal model component set to the main cluster
parameters, and with the maximum temperature and abundance set to the
parameters from fit number 4 in Table~\ref{tab:south_spec_fits}, we
derived a cooling rate of $6.8\pm0.6~M_\odot$~yr$^{-1}$.
Both fits are consistent with a low present cooling rate.
Both fits also had a significantly worse reduced $\chi^2$ than did
either of the first two models presented in
Table~\ref{tab:south_spec_fits}.

We fit a single temperature model to the spectrum of the subcluster
minus the cold core and found a temperature of
$6.3^{+0.6}_{-0.5}$~keV.
If we add a model component for the emission from the main cluster, we
find a subcluster temperature of $5.5^{+0.7}_{-0.6}$~keV.
The former temperature is slightly lower than the value given by
\citet{mfs+98}, but is consistent to within the errors.
Our extraction region is smaller than that used by
\citet{mfs+98} due to the much poorer angular resolution of {\it ASCA}.
Thus, the {\it ASCA} spectrum may have included more emission from the
main cluster, which is hotter.
In any case, the bulk of the subcluster is hotter than might be
expected for a cluster of this mass and X-ray luminosity, which
may indicate that the much of the subcluster gas has been heated by
shocks or adiabatic compression associated with the merger.

\subsection{Temperature and X-ray Surface Brightness Profiles}
\label{sec:south_xray_profiles}

We measured the temperature and surface brightness gradients inside the
subcluster and in front of the cold front (Figure~\ref{fig:south_prof}).
The temperature measurements in front of the cold front were made by
extracting spectra in a wedge of elliptical annuli whose curvature
matched that of the cold front.
The measurements within the subcluster were made from spectra also
accumulated from annular wedges using ellipses self-similar to those in
front of the cold front.
The size and orientation of the wedges inside the subcluster were
determined by the edges of the subcluster, and therefore were not
oriented 180$^{\circ}$ from the wedge in front of the subcluster.
While the annular wedges in front of the subcluster were centered along a
line 15$^{\circ}$ west of north, the wedges inside the subcluster were
centered on a line $\sim 58^{\circ}$ east of south.
The regions used to extract the spectra are shown in
Figure~\ref{fig:south_prof}a.

We fit single-temperature models to these spectra, with the absorption
column set to the Galactic value.
The resulting temperatures from fits to these spectra are shown in
Figure~\ref{fig:south_prof}b.
The gas in front of the cool core is hot, and the temperatures are
consistent (within the large errors) with the temperature in the
main cluster at this radius.
The temperature in the cool core of the subcluster is quite low.
Behind the cool core, the temperatures in the subcluster rise up
to moderately high values.

X-ray surface brightness measurements were made in elliptical annular
wedges with the same shape as those used to extract the spectra, but
with smaller widths.
The resulting surface brightness profile in the 0.3--10 keV band is shown
in Figure~\ref{fig:south_prof}c.
As is clear from the image (Figure~\ref{fig:south_smooth_image}), the highest
surface brightness is associated with the cool core.
There is a very sharp surface brightness discontinuity (a factor $\ga 5$)
at the northern edge of the cool core.
The combination of the surface brightness discontinuity with the low
temperature in the bright region shows that this is a cold front.
The subcluster south of the cool core is much brighter than the gas ahead
of the cold front.

If one assumes that the subcluster is merging for the first time, that the
motion is transonic, and that the mass of the subcluster is much smaller
than that of the main cluster, one would expect the gas far ahead of
the cold front would be undisturbed main cluster ICM.
Also, main cluster emission may be projected in the foreground and
background of the subcluster.
To determine the contribution of undisturbed main cluster emission,
we measured the surface brightness profile of the main cluster in a
wedge to the southeast;
this region is essentially the same region as that
occupied by the subcluster, but reflected across the north-south axis of
symmetry of the main cluster.
We fit a $\beta$-model to the surface brightness at projected radii from
380 to 680 arcsec, which covers the range of radii containing the
subcluster.

The main cluster surface brightness determined from this fit is shown as
a dashed curve in
Figure~\ref{fig:south_prof}c.
More than $\sim$40~kpc ahead of the cold front, the X-ray surface
brightness is consistent with the undisturbed main cluster emission within
the errors.
However, there is some evidence for a rise in the surface brightness just
ahead of the cold front;
the four values at $\sim$0--40~kpc are all slightly higher than expected.
This may indicate that the main cluster gas is compressed ahead of the
subcluster and cold front, by a bow shock and/or by adiabatic compression.

To show more clearly the excess X-ray emission associated with the
subcluster and any compression of main cluster gas ahead of the cold front,
in Figure~\ref{fig:south_prof}d we subtract the fit to the undisturbed main
cluster emission from the surface brightness values in
Figure~\ref{fig:south_prof}c.
In most of the region in front of the subcluster, there are only upper limits
on the excess emission.
There may be some excess emission just ahead of the cold front, but in
the residual surface brightness profile only the point from $\sim$1--10~kpc
appears to be significantly increased.
In the residual profile, the surface brightness of the subcluster is
relatively uniform except for the brighter cool core.
Thus, the apparent fall-off in the surface brightness of the subcluster
with increasing distance from the cold front in Figure~\ref{fig:south_prof}c
may actually be due to projected main cluster emission.

We determined the gas densities in the regions around the cold front
and subcluster by deprojection.
We assumed different geometries for the subcluster and for the gas ahead of
the cold front.
For the densities inside the subcluster, we assumed the subcluster to be
a cone opening up behind the cold front, with an opening angle
determined from the image.
We further assumed that this cone's axis of symmetry lies in the plane
of the sky.
We determined the gas density well ahead of the cold front from the
$\beta$-model fit to the surface brightness of the main cluster discussed
above.
We also assumed a spherically symmetric main cluster to do the deprojection.
We used the deprojected gas densities and the temperatures from spectral
fits to determine the pressures across the cold front.

\section{Hydrodynamical Analysis of Merger}
\label{sec:south_hydro}

We now use the gas temperatures, densities, and pressures derived in
\S~\ref{sec:south_xray_profiles} to analyze the kinematics and hydrodynamics
of the subcluster merger.
Our treatment closely follows that in \citet{vmm01b}.
A schematic view of the geometry of the flow of main cluster gas near
the cold front is shown in Figure~\ref{fig:south_cold_front_diag}),
which is adapted from \citet{vmm01b}.
A bow shock will be present ahead of the cold front if the merger
velocity of the cold front relative to undisturbed main cluster gas
is supersonic.
(There is also a shock in the cold front, but the shocked region will
be very narrow if the cool core is much denser than the main cluster gas.)
If the motion is subsonic, the flow around the cold front is continuous.
For any blunt cold front, there will be a point in front of the
cold front where the velocity of the main cluster gas is zero.
This stagnation point is labeled ``st'' in
Figure~\ref{fig:south_cold_front_diag}).

\subsection{Stagnation Pressure at Cold Front}
\label{sec:south_hydro_stag}

The ratio of the pressures in the main cluster gas far ahead of the cold
front to that at the stagnation point,
combined with temperature measurements of the gas,
can be used to determine the velocity of the cold front
\citep{vmm01b}.
Ideally, we would measure the pressure at the stagnation point in
the hot gas.
However, since the surface brightness of the hot gas at the stagnation
point is actually quite low, we instead measure the pressure in the cool
core just behind the stagnation point where the surface brightness is
much higher.
Because the cold front is a contact discontinuity and the gas is
moving subsonically near the stagnation point, the pressure across the
cold front is expected to be continuous.
The pressure difference between the stagnation point and a point far
upstream
(region~1 in Figure~\ref{fig:south_cold_front_diag})
must be caused by compression of the gas in region~2 by a bow shock
and/or adiabatic compression.

\begin{inlinefigure}
   \ifthenelse{\equal{\figtype}{EPS}}{
     \epsfxsize=.45\textwidth
     \epsfysize=.45\textwidth
     \centerline{\epsfysize=0.83\colwidth\epsfxsize=\colwidth\epsffile{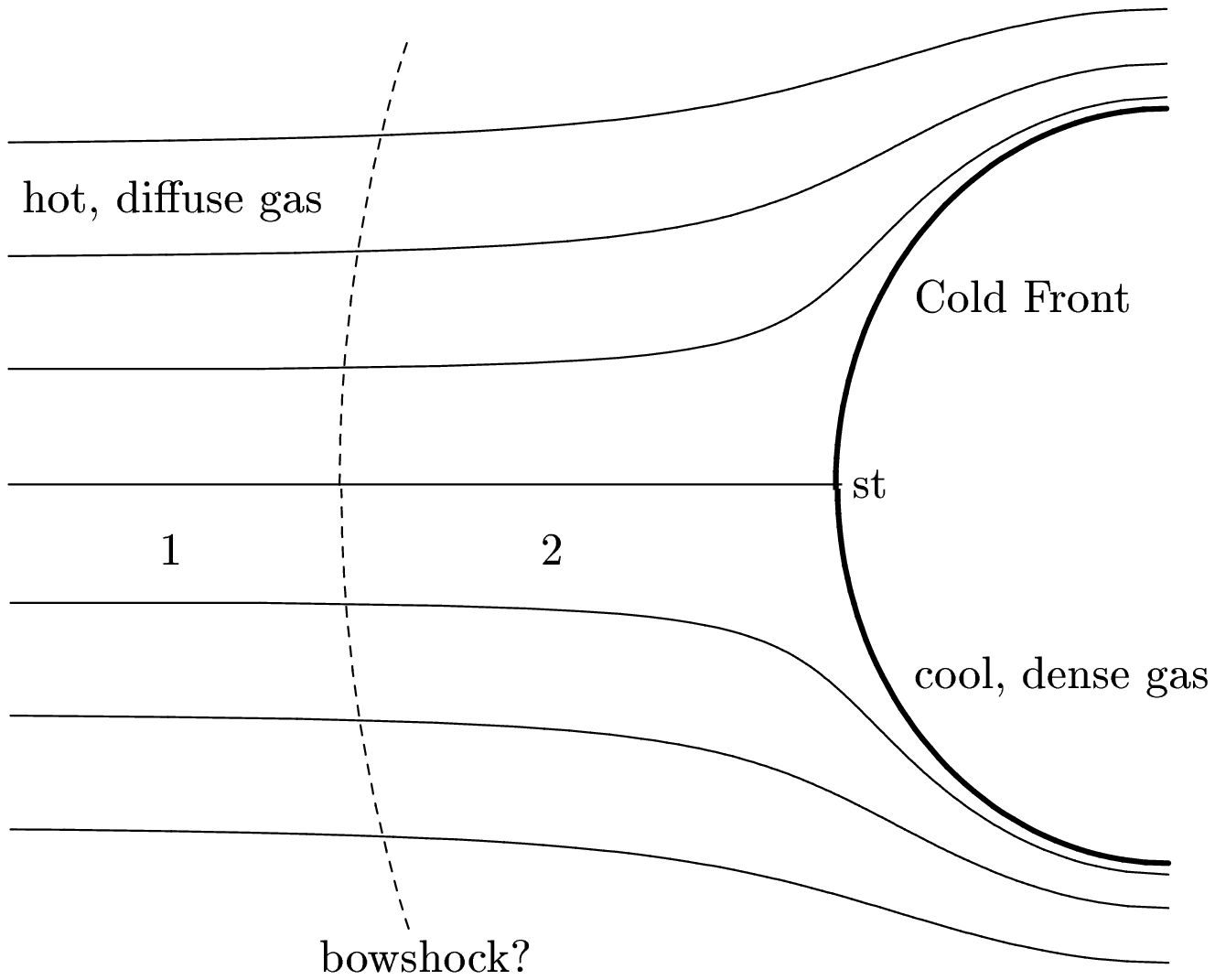}}
  }
  {
  }
\bigskip
\figcaption[f4.eps]{Schematic diagram of the flow of hot main cluster
gas around a blunt cold front.
A bow shock will be present ahead of the cold front if the merger
is supersonic.
The stagnation point is labeled ``st''.
Region~1 is unshocked gas of the main cluster; region~2 is gas which has
passed through the bow shock (if present).
\label{fig:south_cold_front_diag}}
\end{inlinefigure}

The ratio of the pressure at the stagnation point to the pressure in the
far upstream region~1 is given by \citep[e.g.][ \S114]{ll59}
\begin{equation}\label{eqn:south_Pratio}
\frac{P_{\rm st}}{P_1} = \left\{
\begin{array}{cl}
\left( 1 + \frac{\gamma - 1}{2} {\mathcal M}^2
\right)^{\frac{\gamma}{\gamma - 1}} \, , &
{\mathcal M} \le 1 \, , \\
{\mathcal M}^2 \,
\left( \frac{\gamma + 1}{2}
\right)^{\frac{\gamma + 1}{\gamma - 1}} \,
\left( \gamma - \frac{\gamma - 1}{2 {\mathcal M}^2} \,
\right)^{- \frac{1}{\gamma - 1}} \, , &
{\mathcal M} > 1 \, . \\
\end{array}
\right.
\end{equation}
Here $P_{\rm st}$ and $P_1$ are the pressures at the stagnation point and in
region~1, respectively and $\gamma = 5 / 3$ is the adiabatic index for a
fully ionized plasma.
$\mathcal{M} \equiv v_1 / c_{s1}$ is the Mach number of the cold core,
$v_1$ is the cold core's velocity relative to the upstream gas,
and $c_{s1}$ is the sound speed in that gas.

We do indeed measure a higher pressure inside the cold core as
compared to the undisturbed gas in front of the core.
The best fit measurement is $P_{\rm st} / P_1 = 3.4$ which implies a Mach
number of $1.4$.
This measurement assumes that the separation between the main cluster
and the subcluster is equal to their projected separation, so $P_1$ is
measured at the projected radius of the subcluster within the main
cluster.
The formal errors yield a wide range in allowable Mach numbers, from
0 to $3.3$, but since not all the sources of error are independent,
the actual error is probably somewhat smaller.
While a Mach number of zero is allowed, the morphology of the system
makes such a value highly unlikely.
In any case, the best fit value of the pressure ratio requires that the
merger motions be slightly supersonic.

\subsection{Possible Bow Shock}
\label{sec:south_hydro_bow}

For a supersonic cold core, a bow shock should also form in front of the
cold core.
Assuming that the density and temperature of the gas in
region~1 is constant, the bow shock should have a predictable
``stand-off'' distance, $d_s$, which is the shortest distance from
the stagnation point to the bow shock
\citep{vmm01b}.
This distance can be calculated using the approximate method given by
\citet{m49},
and depends only on the value of $\mathcal{M}$ and on the shape of the
cold front.
A useful plot of the stand-off distance versus Mach number is given
in \citet[][Figure 4]{s02}.
For $\mathcal{M} \gtrsim 2$, the stand-off distance of the shock is
not very sensitive to the value of $\mathcal{M}$, while for smaller
Mach numbers the distance increases rapidly.
For the best fit value of $\mathcal{M} = 1.4$, the expected stand-off
distance is $d_s \sim 18$~kpc if we treat the cold front as
a spherical surface with a radius of curvature of 19~kpc.
Such a bow shock would compress the gas in region~2, and should produce a
measurable increase in the X-ray surface brightness, $I_X$, in that
region.
A possible surface brightness excess is seen in the $\sim$20~kpc immediately
upstream from the cold front, but it is significant at only slightly
greater than the $1.7\sigma$ level.
While this surface brightness excess is not visible in the image of the
cluster, we might expect such a feature to be more visible in profile
given the significant azimuthal averaging done to create the profile.

Given enough source photons, the spatial resolution of {\it Chandra}
would be sufficient to measure the expected stand-off distance of the
bow shock, which corresponds to $\sim$5--15\arcsec.
However, the surface brightness in the hot gas ahead of the cold front
is too low to allow $d_s$ to be accurately determined from the available
data.
Thus, all we can conclude is that the expected values of the stand-off
distance are consistent with the (marginal) evidence for an increase
in the X-ray surface brightness within $\sim$20~kpc of the cold front.

We can also use the Rankine-Hugoniot shock jump conditions at the
putative bow shock to independently determine the Mach number
\citep[e.g.][ \S85]{ll59}.
The shock jump conditions yield
\begin{equation}\label{eqn:south_RHjump}
\frac{1}{C} = \frac{2}{\gamma+1} \frac{1}{\mathcal{M}^2} +
\frac{\gamma-1}{\gamma+1} ,
\end{equation}
where $C \equiv \rho_2 / \rho_1$ is the shock compression.
Because we do not have spectra or temperatures determined on the
scale of the bow shock,
we estimate the shock compression from the small increase in the
surface brightness as $C \approx ( I_{X2} / I_{X1} )^{1/2}$.
The observed surface brightness increase in the first 10~kpc is a factor
of $\sim 1.7$, which implies that $C \sim 1.3$.
This implies $\mathcal{M} \sim 1.2$.
It is likely that the finite resolution with which the surface brightness
contrast was determined and projection effects cause the shock
compression to be underestimated.
Thus, this value for the Mach number is consistent with that
determined from the pressure increase at the stagnation point,
If we assume the Mach number determined by the stagnation condition
$\mathcal{M} \sim 1.4$, the expected shock compression is $C \sim 1.6$.
Projection effects (if the cluster is not moving in the plane of the
sky) would cause us to overestimate the stand-off distance and
underestimate the Mach number.
Projection could also cause us to inaccurately determine the true shape
of the cold front.

\section{Merger Kinematics}
\label{sec:south_kine}

Since it provides a consistent fit to the stagnation pressure, the bow shock
compression, and the bow shock stand-off distance, we will adopt the
merger Mach number of $\mathcal{M} \approx 1.4$.
The sound speed in the upstream gas is $\approx 1540$~km~s$^{-1}$, so a
Mach number of $1.4$ implies a merger velocity of $v \approx 2150$~km~s$^{-1}$.

\subsection{Kinematic Model}
\label{sec:south_kine_model}

We now construct a kinematic model for the merger which is consistent
with the X-ray and optical observations of the main cluster and
subcluster.
Because the errors on our determination of the Mach number are large,
the parameters of this model are not well constrained.
We therefore do not suggest that this model is the only possible
interpretation of the data, but that it is merely a ``toy model'' that
is adequate to explain the data given the best-fit values from the
various hydrodynamic tests.
The model, then, is presented as illustrative rather than interpretive,
using the best-fit parameters from the hydrodynamic analyses as the
primary constraints of the model.
We also discuss the implications of this model for the merger.

The parameters of the model are shown in Figure~\ref{fig:model}.
We will approximate the subcluster as a point mass with a single velocity
relative to the main cluster.
We will also assume that the mass of the subcluster is small relative to
that of the main cluster, so we can treat the subcluster as a test
particle falling into the extended mass distribution of the
main cluster.
We put the center of the main cluster at the center of our coordinate
grid, and using polar coordinates, we define the $x$-axis to be parallel
to North and the $z$-axis to be the line of sight, with the positive
$z$-axis extending away from the observer.
Let the vector $\vec{d}$ be the position the subcluster relative to
the main cluster (direction from the main cluster to the subcluster).
The components of $\vec{d}$ are defined by its magnitude $d$,
the angle to the line of sight $\theta_d$ (from the positive $z$-axis),
and the position angle on the plane of the sky $\phi_d$ measured
counterclockwise from the north.
Similarly, let $\vec{v}$ be the velocity of the subcluster relative
to the main cluster, with magnitude $v$, and direction given by
the angles $\theta_v$ and $\phi_v$.
We also define $\psi$ to be $180^\circ$ minus the angle between
$\vec{d}$ and $\vec{v}$.

On the X-ray image (Figure~\ref{fig:south_image}),
the subcluster center is located at a position angle of
$\phi_d = 194^\circ$ from the main cluster.
The position angle of the direction of motion, $\phi_v$ is less certain,
since the curvature of the cold front suggests $\phi_v \sim -15^\circ$
whereas the body of the subcluster is more consistent with
$\phi_v \sim -58^\circ$.
We will adopt the average value of $\phi_v \approx -36^\circ$.

We can estimate the radial component of the relative velocity of
the subcluster from the optical redshifts of the main cluster and
subcluster.
We will adopt the velocity of the subcluster cD galaxy as representative
of the subcluster;
in any case, it is most closely related to the cold front, which was
used to derive the merger velocity.
\citet{bfh+91}
give the redshift of the subcluster cD galaxy as $z = 0.05633 \pm
0.00012$, while the galaxies within $5\farcm25$ of the center of
the cluster have a mean redshift of $0.0538 \pm 0.0050$
\citep{dfl+98}.
Combining the line-of-sight velocity determined from the optical
redshifts with the velocity merger velocity determined in
\S~\ref{sec:south_hydro} from the X-ray data, we find that the direction
of motion of the subcluster relative  to the main cluster is between
$\theta_v = 63^{\circ}$ and $76^{\circ}$ from the line of sight, or
$14^{\circ}$ to $27^{\circ}$ from the plane of the sky.
We adopt the average value of $\theta_v \approx 71^\circ$.

\begin{inlinefigure}
   \ifthenelse{\equal{\figtype}{EPS}}{
     \epsfxsize=.45\textwidth
     \epsfysize=.45\textwidth
     \centerline{\epsfysize=0.93\colwidth\epsfxsize=\colwidth\epsffile{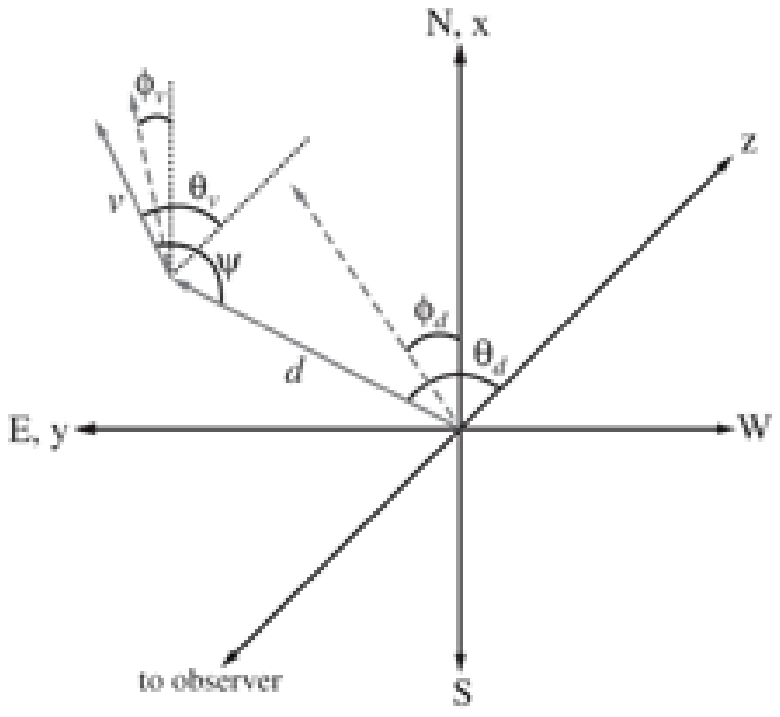}}
  }
  {
  }
\figcaption[f5.eps]{Schematic diagram of the kinematic model.
The position of the subcluster relative to the main cluster is
given by the vector $\vec{d}$, and $\vec{v}$ is the velocity of the
subcluster relative to the center of the main cluster.
The $z$-axis is along the line-of-sight.
The dashed lines are projections of the direction and velocity vectors
onto the plane of the sky.
The dotted line is parallel to the N-axis.
The small inset shows the projection of the direction and velocity
vectors onto the N-$z$ plane, with the dotted line parallel to the
$z$-axis.
\label{fig:model}}
\end{inlinefigure}

It is difficult to determine the angle between the separation of the
main cluster and subcluster, $\vec{d}$, and the line-of-sight.
Initially, we will assume that the main cluster and subcluster
are separated in the plane of the sky ($\theta_d = 90^\circ$),
so that the separation between the main cluster and cluster
is equal to the projected separation,
$d = 730$~kpc.
The angle between $\vec{d}$ and $\vec{v}$ is then $\psi = 53^\circ$,
and the impact parameter of the merger is $b \approx 580$~kpc.

\subsection{Infall Velocity}
\label{sec:south_kine_infall}

We now compare the merger velocity inferred from the hydrodynamics of the
cold front and subcluster with the infall velocity expected for the
subcluster and main cluster.
\citet{rb02} derive a virial mass for the main cluster of $M_{200} =
1.080 \times 10^{15}~M_\odot$ for a virial radius $r_{200} = 2.66$~Mpc.
We will assume that the main cluster mass is much larger than that of
the south subcluster.
We will assume that the subcluster has fallen into the main cluster from
its turn-around distance of 5.5~Mpc,
which is the value if the age of the Universe is 13~Gyr.
As noted above, we initially assume that the main cluster and subcluster
are separated in the plane of the sky.

We will consider two models for the mass distribution and potential of the
main cluster.
We first consider a singular isothermal sphere out to the virial radius,
which has a potential given by
\begin{equation}
\Phi (r) = - 2 \sigma^2 \times \left\{
\begin{array}{cl}
1 + \ln ( r_{200} / r ) \, , & r \le r_{200} \, , \\
r_{200} / r \, , & r \ge r_{200} \, . \\
\end{array}
\right.
\label{eqn:south_poteniso}
\end{equation}
Here, $\sigma = ( G M_{200} / 2 r_{200})^{1/2} \approx 934$~km~s$^{-1}$
is the velocity dispersion.
The infall velocity at the subcluster's current projected distance of
730~kpc is 2520~km~s$^{-1}$.
We also consider a model in which the density within the virial radius is
given by the Navarro, Frenk, \& White model
\citeyearpar[][hereafter NFW]{nfw97}, for which the potential is
\begin{equation}
\Phi (r) = - \frac{G M_{200}}{r_s} \times \left\{
\begin{array}{cl}
\frac{\frac{\ln ( 1 + x )}{x} - \frac{1}{1+c}}
{\ln ( 1 + c ) - \frac{c}{1+c}} \, , & r \le r_{200} \, , \\
\frac{1}{x} \, , & r \ge r_{200} \, . \\
\end{array}
\right.
\label{eqn:south_potennfw}
\end{equation}
Here, $r_s$ is the scale radius, and $x \equiv r/r_s$.
We adopt a concentration parameter $c \equiv r_{200}/r_s = 10$, which is
consistent with NFW's simulations for cluster-mass halos.
For this potential, the predicted infall velocity at the projected
distance is 2740~km~s$^{-1}$.

For either potential, the infall velocity at the projected separation is
somewhat larger than the velocity we determined from the X-ray observations
of the merger hydrodynamics.
Given the large errors in the determinations of the velocities, 
this difference may not be significant.
As first noted by \citet{msv99}, the degree of agreement between
the merger velocity determined by hydrodynamical measurements
and that expected from infall can be used to test the hypothesis that
the intracluster medium is predominantly a non-relativistic, thermal
plasma.
That is, the calculation of of the merger velocity from shock conditions
(equation~\ref{eqn:south_RHjump}) assumes that the merger shock energy
is thermalized, and is not converted into relativistic particles, or
magnetic fields, or turbulence.
All of the hydrodynamic diagnostics require that the intracluster
medium act as a $\gamma = 5/3$ gas.
Thus, the difference between our hydrodynamical estimate of the merger
velocity and the predicted infall velocity (assuming the main cluster and
subcluster are separated in the plane of the sky) could suggest
that the kinetic energy of the merger is not thermalized particularly
efficiently, but instead goes partially into turbulence, magnetic fields, or
relativistic particles.
As noted by \citet{msv99}, this argument is somewhat circular, as the
mass of the cluster was also determined from hydrostatic equilibrium
assuming purely thermal pressure support.
Given the uncertainties in the determination of merger velocity and infall
velocity, we will instead take the crude agreement between the two
speeds as an indication that at least a significant fraction of the
merger energy ($\ga 50$\%) is thermalized.

Our initial estimate of the infall velocity was based on the
assumption that the main cluster and the subcluster were separated in the
plane of the sky ($\theta_d = 90^\circ$).
If this is not true, the actual separation $d$ will be larger than the
projected separation of 730 kpc, and the predicted infall velocity will be
lowered.
To illustrate this effect, we construct a consistent model for the merger
geometry and kinematics in which the merger velocity equals the predicted
infall velocity.
As we move the subcluster further out in the main cluster potential, the
density and hence the pressure in the ambient medium drops, thereby
increasing the pressure ratio used to determine the Mach number.
This in turn increases the Mach number we would measure, lessening the
need to place the subcluster significantly in front of or behind the
main cluster.

Given the large errors in the two numbers, our model is certainly not a
unique solution, but is consistent with the current best-fit values
of the parameters.
For the isothermal potential
(equation~\ref{eqn:south_poteniso}), this consistent solution requires that
$d \approx 820$~kpc, while for the NFW potential
(equation~\ref{eqn:south_potennfw}) we find $d \approx 860$~kpc.
The Mach numbers we derive are $\mathcal{M} = 1.6$ for the isothermal
potential and $\mathcal{M} = 1.7$ for the NFW potential.
The corresponding velocities are $v \approx 2460$~km~s$^{-1}$ and $v
\approx 2610$~km~s$^{-1}$ respectively.
For both solutions, we assume that the sound speed does not vary over
the range of radii in question.
If we were to allow the sound speed to vary, it would decrease slightly
at larger radii, increasing the physical separation that we determine.
To be specific, we will adopt the isothermal result.
This distance implies that the angle between the separation and the
line of slight is either $\theta_d \approx 64^\circ$ or
$\theta_d \approx 116^\circ$.
The former value implies that the $\psi \approx 66^\circ$, which means
that the subcluster is moving nearly perpendicular to the radius from
the center of the main cluster, and about to start exiting the cluster.
The observed morphology of the X-ray image of the subcluster seems
inconsistent with this interpretation.
Thus, we adopt the solution with $\theta_d \approx 144^\circ$ and
$\psi \approx 46^\circ$, in which the subcluster is moving into the
main cluster, probably for the first time.
In summary, our consistent kinematic model has
$v \approx 2460$~km~s$^{-1}$,
$\theta_v \approx 71^\circ$,
$\phi_v  \approx -36^\circ$,
$d \approx 820$~kpc,
$\theta_d \approx 116^\circ$,
$\phi_d \approx 194^\circ$,
and
$\psi \approx 46^\circ$.
The subcluster is closer to us than the main cluster, and is falling
into the main cluster.

\subsection{Angular Momentum and Impact Parameter}
\label{sec:south_kine_lambda}

As noted above in \S~\ref{sec:south_xray_image},
the direction of the merger velocity is not parallel to the separation of
the centers of the main cluster and subcluster.
The transverse component of the merger velocity lies at an
angle of $\sim$50$^\circ$ with respect to the projected separation
of the two clusters.
This implies that this is an offset merger; the collision is occurring
with a nonzero impact parameter and angular momentum.
If we adopt the consistent model for the merger kinematics which we have just
discussed, the angle between the velocity and the separation is
$\psi \approx 46^\circ$, and the impact parameter for the collision is
$b = d \sin \psi \approx 750$~kpc.
This is $3.7$ times the core radius we determine from a $\beta$-model
fit to the cluster.

A useful dimensionless form for the angular momentum is given by
the $\lambda$ parameter, defined as
\citep{p69}
\begin{equation}
\label{eqn:south_lambda}
\lambda \equiv {J|E|^{1/2}\over GM^{5/2}}\ .
\end{equation}
Here $J$ is the total angular momentum of the merged cluster,
$E$ is its total energy,
and $M$ is its mass.
We estimated the value of $\lambda$ implied by the merger velocity and
impact parameter of the subcluster
by differentiating equation~(\ref{eqn:south_lambda}),
assuming the mass of the subcluster is much smaller
than that of the main cluster.
We assumed that the orbital angular momentum of the subcluster was
parallel to the initial angular momentum of the main cluster, and
ignored the initial internal angular momentum of the subcluster.
Using the kinematic parameters for our consistent model for the merger, we find
$\lambda \approx 0.21$.
Fixing $v$, $\theta_v$, and the projected angles $\phi_d$ and $\phi_v$
at the values from the consistent model, and allowing the value of
$\theta_d$ to vary, we find a minimum value of $\lambda$ of about 0.16.
If the initial spin of the main cluster is not aligned with the orbital
angular momentum of the subcluster, the value of $\lambda$ would be
an upper limit.
These values are both somewhat larger than the median
values of 0.05--0.1 expected from tidal effects in large scale
structure.
This may reflect the large uncertainties in the kinematic parameters.
If this large angular momentum is correct,
it might be the result of tidal effects associated
with the triple merger occurring in Abell~85 (i.e., the fact that there
are two merging subclusters).
Alternatively, it may be that mergers with small subcluster have a
larger range of values of $\lambda$, which average out when many
small subclusters merge to form a larger halo.

\section{Suppression of Conduction Across the Cold Front}
\label{sec:south_trans_conduction}

As discussed in \S~\ref{sec:south_xray_image}, the cold front is seen as
a sharp surface brightness discontinuity, with a dramatic increase in
surface brightness in the cold gas over only a few kiloparsecs.
 From the surface brightness profile in Figure~\ref{fig:south_prof}, the
change in surface brightness occurs over at most 20~kpc---half the
width of the elliptical region on the cold core or the width of about 2
bins ahead of the cold front.
The raw image suggests that it is actually quite a bit narrower than
this, but we are limited by photon statistics to the aforementioned
resolution.
Unfortunately, the same photon statistics prevent us from measuring the
temperature gradient on this size scale.
However, because the pressure is continuous on the large scale of our
measurements and is presumed to be continuousl on the smaller scale of
the cold front as wel, the observed density gradient should be
accompanied by a temperature gradient of the opposite sign and with the
same length scale.
Therefore, while the width of the gradient is determined from the
surface brightness, we can assume that the same width applies to the
temperature gradient.
Thermal conduction should smear out any such sharp edges to a length a
few times the electron mean free path in a relatively short time.
However, conduction appears to be suppressed in the case of Abell~85, as
it does in other clusters with observed cold fronts
\citetext{e.g. Abell~2142, Abell~3667; \citealp{ef00,vmm01b}}.

\citet{ef00} showed that thermal conduction should smear out the
temperature gradient to a width $\delta r$ on a characteristic timescale
\begin{equation}
\delta\tau = \frac{\delta r}{\bar{v}} ,
\end{equation}
where
\begin{equation}
\bar{v} = \frac{2}{3} \frac{\kappa}{n_e k_B T_e} \frac{{\rm d}(k_B T_e)}{{\rm d}r}
\end{equation}
is the characteristic velocity of the diffusion.
Here,
\begin{equation}
\kappa = 8.2\times10^{20}
\left( \frac{k_B T_e}{10 \, {\rm keV}} \right)^{5/2}
\, {\rm erg} \, {\rm s}^{-1} \, {\rm cm}^{-1} \, {\rm keV}^{-1}
\end{equation}
is the thermal conductivity, and $n_e$ and $T_e$ are the electron number
density and temperature, respectively.
For the upper limit on the width of the cold front of 20~kpc, the
diffusion timescale is $2.0\times10^6$~yr.

At the subcluster's current velocity and distance from the cluster
center, the relevant timescale for interaction is roughly
$d / v \approx  5.6 \times 10^8$ yr.
This means that in order for thermal conduction to have failed to erase
the sharp edge of the cold front, conduction must be suppressed by at
least a factor of 280--2700.
In fact, since the rate of conduction is independent of density, the
time over which conduction has had a chance to act is probably somewhat
longer, meaning that the degree of suppression is probably even higher.

One mechanism that has been suggested for this suppression is the
existence of a magnetic field perpendicular to the direction of
diffusion, i.e. parallel to the surface of the discontinuity
\citep{vmm01b}.
A tangled magnetic field would serve the same purpose, with a maximum
loop size equal to the width of the front.

\section{Subcluster/Radio Source Interaction}
\label{sec:south_radio}

The possible existence of a large scale radio halo or relic in the
southern subcluster in Abell~85 was raised by
\citet{bpl98}.
As shown in Figure~\ref{fig:south_radio},
the diffuse emission was later resolved by
\citet{gf00}
into what may be a tailed source associated with a dead or dying AGN in
the dominant galaxy of the subcluster, where the nuclear emission is
very faint compared to the radio lobes.
If this is in fact its origin, the shape of the tail is well explained
by the merger interaction.
Figure~\ref{fig:south_radio} shows that the source associated with
the cD galaxy in the subcluster (Source D) has possible weak nuclear
emission, and a lumpy C-shaped extended source to the southeast of
the cD galaxy.
The radio emission has no significant extent north of the subcluster's
leading edge, and the bulk of the radio emission follows the bright
X-ray arc through the subcluster.
If it is indeed a tailed galaxy, this shape implies that the same ram
pressure forces which have shaped the subcluster have also bent the
lobes of the AGN into their current shape.

An alternative explanation for the diffuse radio emission is that this
is a small, merger-induced halo in the subcluster.
The fact that its surface brightness correlates roughly with that of the
X-ray gas is consistent with findings for other halos, as is its steep
spectral index
\citep[$\alpha^{1.4}_{0.3} \sim 2\!\!-\!\!2.5$;][]{gf00}.
On the other hand, the radio source is quite overluminous compared to
the expected radio power derived from the empirical relation between
X-ray temperature or luminosity and radio halo power
\citep{lhb+00,f00}.
For the observed X-ray luminosity of the subcluster of $5 \times
10^{43}$~erg~s$^{-1}$ (rest frame 0.1--2.4~keV), the empirical relation
would require a monochromatic radio power of only $4 \times
10^{21}$~W~Hz$^{-1}$ at 1400~MHz in the cluster rest frame, compared to
the observed value of $1.8 \times 10^{24}$~W~Hz$^{-1}$
\citep{f01}
Furthermore, halos have previously only been observed in the hottest and
most massive clusters, never in a cluster as small and cool as this
subcluster.
While previous observations may have been biased towards finding halos
in hot clusters
\citep{ks01},
it would nonetheless be surprising to find a halo in such a cool, low
luminosity cluster.
Our observations are therefore more consistent with the interpretation
of the source being a tailed radio galaxy
\citep{gf00}
than with it being a radio halo or relic.

Source ``B'' in Figure~\ref{fig:south_radio} is a narrow angle tail
(NAT) source studied in detail by \citet{oo85}.
It has a redshift of $z=0.0579$ \citep{wcs+99}, which is within the
dispersion of the main cluster.
Since the line of sight component of the subcluster's infall velocity is
so small, however, it is not possible to determine whether or not this
galaxy is a member of the subcluster or of the main cluster
based solely on radial velocity information.
The direction of the tail, plus the fact that the radio tail does not
appear to be interacting with the flow around the subcluster, suggest
that it is seen in projection in front of or behind the subcluster,
and that it is a member of the main cluster.

\begin{inlinefigure}
   \ifthenelse{\equal{\figtype}{EPS}}{
     \epsfxsize=.45\textwidth
     \epsfysize=.45\textwidth
     \centerline{\epsfysize=0.86\colwidth\epsfxsize=\colwidth\epsffile{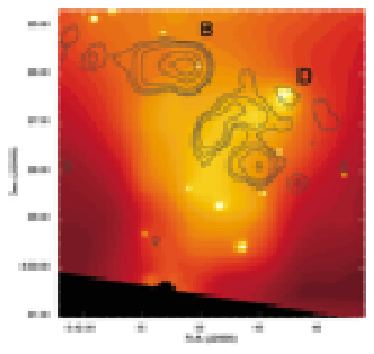}}
  }
  {
  }
\figcaption[f6.eps]{Radio contour map at 90~cm of the ``B'' and ``D''
sources
\protect\citep[from][]{gf00}
in the region of the southern subcluster, overlaid on the
adaptively-smoothed {\it Chandra} image of the subcluster (color, from
Figure~\ref{fig:south_smooth_image}).
The D source appears to be associated with the cD galaxy in the south
subcluster, while the B source is associated with another cluster galaxy.
\label{fig:south_radio}}
\end{inlinefigure}

\section{Summary}
\label{sec:south_summary}

Our analysis of the south subcluster in Abell~85 from $\sim 37$~ksec of
{\it Chandra} data has revealed several interesting features.
The most obvious is a confirmation that the subcluster is indeed merging
with the the main cluster.
The subcluster contains a remnant cold core which has survived the early
stages of the merger.
It is smaller and more discrete than similar structures found in other
clusters such as
Abell~2142
\citep{mpn+00},
Abell~3667
\citep*{vmm01b},
and RX~J1720.1+2638
\citep{mmv+01},
and is perhaps more akin to the ``bullet'' in 1E0657-56 \citep{mgd+02}
or the ``tongue'' seen in Abell~133 \citep{fsk+02}.

Based on the ratio of the pressure at the stagnation point of the cold front
to that far upstream, on the standoff distance of a possible bow shock,
and on the shock compression from the bow shock, we find a consistent
Mach number and velocity for the merger of
$\mathcal{M} \approx 1.4$ and
$v \approx 2150$~km~s$^{-1}$.
By comparing this velocity to the radial velocity of the subcluster
relative to that of the main cluster, we have determined that the merger
velocity is about
$19^\circ$ from the plane of the sky.
We find a consistent kinematic model for the merger in which the
subcluster is in front of and falling into the main cluster.
This model is consistent with the expected merger velocity if the
subcluster and main cluster have fallen towards one another due to
gravity from their turn-around distance in the Hubble flow.

The X-ray observations indicate that this is an offset merger with
a finite impact parameter and a significant angular momentum.
A crude estimate based on our consistent kinematic model suggests
an angular momentum parameter of $\lambda \sim 0.2$, which is somewhat
larger than the median values expected due to tidal torques.

Magnetic fields in the cold core may be responsible for suppressing
thermal conduction across the cold front.
This would explain the sharpness of the front, which should be smeared
out by conduction in the absence of a magnetic field.
The magnetic fields in the cold core may be high as a result of a
cooling flow or the AGN located in the central cD galaxy.

We confirm the assertion that the diffuse radio structure
in the subcluster is not a cluster radio halo or relic, but is more likely
to be a tailed galaxy with a weak or dead nucleus.
We also show that its morphology has been shaped by ram pressure in the
merger interaction.

\acknowledgements
Support for this work was provided by the National Aeronautics and Space
Administration, primarily through {\it Chandra} Award Number
GO0-1173X,
but also through
GO1-2122X,
and
GO1-2123X,
all issued by the {\it Chandra} X-ray Observatory Center, which is
operated by the Smithsonian Astrophysical Observatory for and on behalf
of NASA under contract NAS8-39073.
We are grateful to Gabriele Giovannini and Luigina Feretti for
providing detailed information about and images of the radio sources.
We also thank Maxim Markevitch for helpful discussions regarding the
kinematic model.

\end{document}